\documentclass[twocolumn,prl]{revtex4}

\usepackage{graphics}

\usepackage{epsfig} %Include figure files

\usepackage{dcolumn}% Align table columns on decimal point

\usepackage{amsmath}

\begin{document}

\title{Modulation theory of quantum tunneling into a Calogero-Sutherland fluid.}
\author{D.B. Gutman}
\affiliation{$^1$Department of Physics, University of Florida,
Gainesville, FL 32611,US}

\date{\today}

\begin{abstract}
Quantum hydrodynamics of interacting electrons with a parabolic single particle spectrum 
is studied using the Calogero-Sutherland model.
The effective action and modulation equations, describing evolution of  
periodic excitations in the fluid, are derived.
Applications to the problem of a single electron tunneling into the FQHE edge state are discussed.
\end{abstract}

\maketitle
It is known since 60's\cite{Luttinger} that Landau Fermi liquid theory is not applicable to
one dimensional metals.
The thermodynamic of one dimensional electronic system is explained by 
Tomonaga-Luttinger (TL) model\cite{Stone-book,Tsvelik}.
This model is a simplified version of a quantum hydrodynamics
that takes into account only a single mode of sound  excitation (phonons)
of an  electronic fluid. This approximation is equivalent to  replacing   
a  true electronic single-particle spectrum by the linear spectrum of  Dirac's fermions.
Though the TL model accounts for thermodynamics it fails  
for effects that involve breaking of the  particle-hole symmetry.
In other words only the diagonal part of kinetic coefficient matrix\cite{Landau5} can be found using 
this approach, while all non-diagonal terms vanish in this approximation.

Among phenomena missed by the TL approximation are
thermopower, photovoltaic effect and  Coulomb drag with a small momentum transfer.
To solve any of this problems one needs to use quantum hydrodynamic without making
one mode approximation.

The problem of single particle  spectrum curvature and electron-electron interaction was recently 
addressed in Ref.\cite{Khodas}, who proposed to 
mimics  a nonlinear  spectrum by  two types of electrons with different Dirac's spectra.
This approach agrees with perturbative expansion of exact results for Calogero-Sutherland (CS) model\cite{Pustilnik2}.

Here, we would follow a different route  of "exact" bosonization, 
originally  developed in Ref.\cite{JevickiSakita} for the sake of matrix models.
As expected\cite{Haldane}, a curvature of fermionic spectrum leads 
to the  cubic in density  terms in a bosonic Hamiltonian.
Though formally a bosonic description  is achieved, the resulting theory 
is non linear.  This renders this method extremely difficult for applications 
and no new results based on it are reported, as yet.

To gain an understanding of one dimensional hydrodynamics we study the special case
of electrons with singular $1/r^2$ 
interaction. This problem is known as the Calogero-Sutherland model  (CS)\cite{Calogero} and has
numerous applications in condensed matter and nuclear physics\cite{reviews}. 
Being exactly solvable\cite{Forrester,Haldane-Ha}   CS model  posses an infinite number of integrals of motion.
That render its hydrodynamics special and easier to analyze.

The connection between the hydrodynamic excitations of the CS fluid
and correlation functions of various operators is not straightforward.
The simplest known example is the soliton, a solution of hydrodynamical equation of motion propagating without changing its shape.  In original description  it corresponds to the many body excitation, called  anyon. 
Anyons  obey   a special "fractional" statistics, 
which is neither Fermi nor Bose like.
The correlation function of anyons is similar to the one of free fermions,
with the values of Fermi velocity and momentum renormalized by interaction\cite{Gutman}.

In order to find other correlation function (such as electrons' Green function) one needs  to solve 
the  system of saddle point equations, that  are  the  continuity and the Euler equation
for CS fluid. This is a very difficult problem.  
It is considerably simplified if the solution is a periodic wave with slowly changing wave parameters.
In this case, the  original hydrodynamic equations may be replaced 
by Whitham's modulation equations\cite{Whitham}.
To put it simple, the TL model  describes particle-hole excitations of fermionic problem  
as sound modes (vibrations)  of one dimensional harmonic string. The modulation theory describes 
particle-hole excitations as vibration of an {\it anharmonic} string.

The modulation technique is well developed for various models of 
dissipationless hydrodynamics.
In particular for the Benjamin-Ono equation\cite{Matsuno,Matsuno-book},
that describes dynamics of internal waves in stratified fluids of great depth.  
As it was stressed recently\cite{Abanov}, 
this two problems are mathematically related. Remarkably, the classical BO equation appears in FQHE\cite{Eldad}, governing the evolution of a semiclassical wave packets containing a large number of fermions.
It is therefore not surprising  that some formal aspects of the derivations
of modulation theory for the BO and the CS models are similar.
The final results of these two models are nevertheless different.

The structure of this work is a following:
starting with the microscopic description of the CS model we pass into its
hydrodynamics (the technical details are given in the Appendix).
Next, we consider periodic density excitation of the CS model and derive its evolution 
equations. Finally, we apply this machinery to the problem of a single electron tunneling into the CS liquid.

The microscopic Hamiltonian of CS model is given by 
\begin{equation}\label{Hamiltonian}
H=-\frac{\hbar^2}{2m}\sum_{i=1}^{2N} \partial^2_i +
\left(\frac{\pi}{L}\right)^2\sum_{i>
j}^{2N}\frac{\lambda(\lambda-1)}{\sin^2\left(\frac{\pi}{L}
(x_i-x_j)\right)} \, .
\end{equation}
Here $\lambda$ is a strength of particle interaction, $m$ is an electron mass, $x_i$ 
is a coordinate of the particle on a circle with the perimeter $L$.
Now on we use the  convention $\hbar=1, m=1$.

On large scales and long time CS model is described by hydrodynamical Hamiltonian 
\begin{equation}\label{e5}
H=\int dx\bigg[ \frac{1}{2}\rho v^2+U[\rho]\bigg] \, 
\end{equation}
(the details of derivation that follows  Ref.\cite{JevickiSakita,Awata} are presented in Appendix). 
The whole specific of CS model is incorporated in potential energy term
\begin{equation}\label{e11}
U=\frac{\pi^2\lambda^2}{6}\rho^3-\frac{\pi\lambda(\lambda-1)}{2}\rho^H\rho_x+\frac{(\lambda-1)^2}{8}\frac{(\rho_x)^2}{\rho} \, ,
\end{equation}
where the  Hilbert transform is defined as 
\begin{equation}
\rho^H(x)=\frac{1}{\pi} P \int dx' \frac{\rho(x')}{x'-x} \, .
\end{equation}
The conventional TL model can be obtained by expansion of  the Hamiltonian (\ref{e5})
up to second order in fluctuations (due to conservation laws the velocity $v$ and
density operators $\delta \rho$ are  of the same order) and neglecting the 
high gradient terms.

It is convenient to reformulate the problem in Lagrangian description,  
defining the action 
\begin{equation}
\label{action}
S[\rho,v]=\int dxdt\bigg[-v\partial_x^{-1}\partial_t\rho-\frac{1}{2}\rho v^2-U[\rho]\bigg] \, .
\end{equation}
This  action reaches its minimum provided saddle point equations are satisfied:
the continuity 
\begin{equation}\label{e11d}
\rho_t+\partial_x(\rho
v)=0
\end{equation}
and Euler equation
\begin{equation}\label{e12}
v_t+vv_x+w_x=0 \,. 
\end{equation}
Here  the enthalpy 
\begin{equation}\label{e12}
w=\left(\frac{\delta U}{\delta \rho}\right)= \frac{\pi^2}{2}\rho^2+
\pi \rho_x^H-\frac{1}{4}\partial_x\left(\frac{\rho_x}{\rho}\right)-
\frac{1}{8}\left(\frac{\rho_x}{\rho}\right)^2.
\end{equation}
The dependence on the interaction constant in the limit of strong interaction
($\lambda \gg 1$) had been  eliminated by the rescaling
($\nolinebreak{\tilde{x}=x/\lambda, \tilde\rho=\rho\lambda,
\tilde{t}=t/\lambda}$). Now on we use the rescaled coordinates, omitting the tilde.

To derive modulation equation it is convenient to integrate 
the velocity field out
\begin{equation}
L=
\int dx  
\left(
\frac{(\partial_x^{-1}\partial_t\rho)^2}{2\rho}-
\frac{\pi^2}{6}\rho^3+\frac{\pi}{2}\rho^H\rho_x-\frac{1}{8}\frac{\rho^2_x}{\rho}
\right)\, .
\end{equation}
This Lagrangian has one-periodic density wave solutions \cite{Polychronakos} 
\begin{equation}
\rho_{\rm s}(\theta)=\rho_0+\frac{k}{2\pi}\frac{\sinh(a)}{\cosh(a)-\cos(\theta)}\, ,
\end{equation}
where $\theta= kx-\omega t$.
The  dispersion relation between amplitude of the wave 
\begin{eqnarray}
A=\frac{k}{2\pi}\frac{1}{\sinh(a)}
\end{eqnarray}
and a wave vector is determined by
\begin{equation}
\tanh(a)=\frac{\pi\rho_0k^3}{\omega^2-\pi^2\rho_0^2k^2-\frac{k^4}{4}} \, .
\end{equation}
To proceed further we replace the Polychronakos solution with a modulated one 
\begin{equation}
\theta(x,t)=k(x,t)x- \omega(x,t) t \, 
\end{equation} 
and allow $\rho_0$ to depend on coordinate and time as well.
Strictly speaking this is no longer the minimum of the action (\ref{action}).
However under the condition the modulation technique works, 
this solution minimizes the action "in average".
We define  Lagrangian averaged over one period of oscillations 
\begin{equation}
\bar{L}=\int_0^{2\pi} \frac{d\theta}{2\pi} L[\rho_s(\theta)] \, .
\end{equation}
After some straightforward, though  lengthy calculation, 
one finds a Lagrangian of anharmonic string 
\begin{eqnarray}&&
\hspace{-1cm}2\bar{L}\!=\!\frac{\omega^2}{2\pi k}\!-\omega\rho_0\!+\!\frac{\gamma^2}{\rho_0}\left(\!1\!-\frac{k^2}{2\omega}\right)\!+\gamma k-\!
\frac{k^3}{24\pi}-\frac{\pi^2\rho_0^3}{3}\,.
\end{eqnarray}
Here $\gamma\equiv-\partial_x^{-1}\partial_t\rho_0$.
Applying the  least action condition  to the averaged Lagrangian 
\begin{eqnarray}&&
\label{least_action}
\left(\frac{\partial \bar{L}}{\partial \omega}\right)_t=\left(\frac{\partial \bar{L}}{\partial k}\right)_x \nonumber \\&&
\left(\frac{\partial \bar{L}}{\partial \gamma}\right)_t=\left(\frac{\partial \bar{L}}{\partial \rho_0}\right)_x \, 
\end{eqnarray}
one obtains modulation eqs. for  CS model
\begin{eqnarray}&&
\label{ea1}
\left(\frac{\omega}{\pi k}+\frac{\gamma^2}{2\rho_0}\frac{k^2}{\omega^2}\right)_t+
\left(\frac{\gamma^2}{\rho_0}\frac{k}{\omega}+\frac{\omega^2}{2\pi k^2}+\frac{k^2}{8\pi}\right)_x=0\nonumber  \\&&
\left(\frac{\gamma}{\rho_0}\bigg[1-\frac{k^2}{2\omega}\bigg]\right)_t+\left(\frac{\gamma^2}{2\rho^2_0}\bigg[1-\frac{k^2}{2\omega}\bigg]
+\frac{\pi^2\rho^2_0}{2}\right)_x=0\nonumber \\&&
\partial_t\rho_0+\partial_x\gamma=0\nonumber\\&&
\partial_tk+\partial_x\omega=0 \, .
\end{eqnarray}
These eqs.  govern the  dynamics of the wave parameters in the density excitations 
propagating through CS fluid.

Next,  we apply this theory to study an evolution of a distortion caused by adding one electron to the fluid.
This problem is a prototype for a quantum  {\it tunneling} 
from the normal metal into the edge state of FQHE.  

An added electron  causes a density fluctuation that 
splits into  two chiral parts, moving in the opposite directions.
Using  Riemann invariants\cite{Matytsin} 
\begin{eqnarray}&&
u=v+\pi\rho \\&&
\bar{u}=v-\pi\rho\, 
\end{eqnarray}
one approximates  eqs.(\ref{e11d},\ref{e12}) by
\begin{eqnarray}&&
\label{b2}
u_t+uu_x=0\nonumber  \\&&
\bar{u}_t+\bar{u}\bar{u}_x=0.
\end{eqnarray}
The tunneling at point $x=0$ corresponds to the initial conditions  
$u_0(\xi)=\frac{\epsilon}{\xi^2+\epsilon^2}$,
(where $\epsilon$ has a scale of an electron wave length). 
To study right chiral sector we pass into a reference frame moving with
a sound velocity to the right  
($\nolinebreak{\xi=x-\pi\bar{\rho}t}$).  
Solving Hopf eq.(\ref{b2}) by Godograph method we find an implicit solution
\begin{equation}
\label{Godograph}
u=u_0(\xi-ut)\, .
\end{equation}
Solving cubic equation(\ref{Godograph}), we find an explicit solution
\begin{eqnarray}&&
\label{b3}
u= 
\left\{
\begin{array}{l}
\frac{\xi}{t}, \,\,\, 0<\xi<\xi_- 
\nonumber \\
\nonumber \\
\frac{\epsilon}{\xi^2}, \,\, \xi>\xi_+\, ,
\end{array}
\right.
\end{eqnarray}
where $\nolinebreak{\xi_-\equiv 3(\epsilon t/4)^{1/3}, \xi_+\equiv t/\epsilon}$ are tailing and leading edge coordinates.
Inside  the interval  ($\nolinebreak{\xi_-< \xi< \xi_+}$)  solution of eq.(\ref{Godograph}) is  multivalued
and we denote three different branches  by $f_1>f_2>f_3$.
The multi-validity of solution reflects a wave breaking phenomena.
The single value solution inside the interval ($\xi_-<\xi<\xi_+$)  is restored by keeping 
the second-order spatial derivatives in eq. (\ref{e12}).
An elegant way of dealing with second-order  gradient terms  
is to use the modulation technique developed above. 
Let us assume that for ($\xi_-<\xi<\xi_+$) density  is  
an oscillating function satisfying modulation eqs.(\ref{ea1}) 
with a proper boundary condition\cite{Pitaevsky}.

At the boundaries, the particle density found from Hopf equation, 
should  match the density, averaged over the period of fluctuations,
inside  the oscillating interval
\begin{eqnarray}&&
\label{b4}
\langle \rho\rangle=\rho_0+\frac{k}{2\pi}. 
\end{eqnarray}

%\begin{eqnarray}&&
%\label{modulation_eq}
%c_t+\left(\frac{c^2}{2}+
%\frac{k^2}{8}\right)_x=0 \\&&
%V_t+\frac{1}{2}\left(V^2+\pi^2\rho_0^2\right)_x=0\\&&
%{\rho_0}_t+(V\rho_0)_x=0\\&&
%k_t+(ck)_x=0
%\end{eqnarray}
In addition, the  amplitude of the oscillation vanished  at the trailing edge 
\begin{eqnarray}&&
\label{b4}
A=0, \,\,\xi=\xi_- \,.
\end{eqnarray}
and  the wave vector vanishes at the leading edge  
\begin{eqnarray}
k=0,\,\xi=\xi_+\,.
\end{eqnarray}
It is convenient to define  phase velocity $c=\omega/k$ and  hydrodynamic velocity $V=\gamma/\rho_0$.
In a new variables eq.(\ref{b4}) can be rewritten as 
\begin{eqnarray}
c=s+\frac{k}{2}\,.
\end{eqnarray}
Solving eqs.(\ref{ea1}), we find  that electron tunneling into CS liquid excites density wave with 
phase velocity and wave vector given by
\begin{eqnarray}&&
\label{b1}
c=\frac{f_1}{2}\simeq\frac{\xi}{2t}\nonumber \\&&
k=f_1-f_2\simeq\frac{2\epsilon}{t}\sqrt{\frac{t}{\epsilon \xi}-1}\nonumber\\&&
\rho_0=\bar{\rho}+\frac{f_3}{2\pi}\simeq\bar{\rho}+\frac{\epsilon^2}{2\pi\xi}\,\,.
\end{eqnarray} 
As we see, the excitation that follow electron tunneling are quite different from the anyon tunneling\cite{Gutman}.
The later  resulted in the creation of a single soliton that was propagating
 with a constant velocity  through the fluid. Electron tunneling causes a spreading density evolution 
 that consist many picks and linearly increasing phase velocity.
 
The large number of oscillations experienced by the fluid density and the 
smooth dependence of wave parameters a posteriori justifies the validity of modulation technique.
One can check, using eq.(\ref{b4}), that solution eq.(\ref{b1}) respects the particle conservation and 
accounts for the half  electrons moving to the right
(solution for another half of electron moving to the left differs 
by sign  change of a sound velocity). 

The  periodic density wave that develops after the tunneling  can be viewed  as a superposition of anyons,
with individual anyone corresponding to the picks of the density oscillations.
Therefore  eq.(\ref{b1}) describe a {\it decay } of an electron
into a large number of quasiparticles of CS model.
This process may be detected  by measurement the time dependence of  a current 
that follows the tunneling event.
Unlike the fractional charge measurements in shot noise 
experiments\cite{Reznikov, Glattli}, the tunneling discussed above is from normal metal to the FQHE state.
Therefore the fractionalization of the elementary charge does not show up  in a 
low frequency shot noise, but in the finite frequency noise of a  current pulse.

In this work we developed  modulation theory for the  CS model.
We applied this theory to the problem of a single-electron  tunneling. 
We found an evolution of a current pulse that followed the tunneling event.

This work was motivated by discussion with I. Gruzberg and P.B. Wiegmann, from whom I learned
various mathematical ideas used in this paper.
 I benefited from discussion  with A. Kamenev, D. Maslov, A. Mirlin, M.Stepanov and  M.Stone.
I acknowledge the Memorial University of Newfoundland, where large part of this work was done, 
for the hospitality. 
My research was supported by  NSF-DMR-0308377. 
\section{Appendix:Derivation of Hydrodynamical Theory}
Consider a ring geometry.
Electron's coordinate is represented by a complex variable
$z_n=Le^{i\theta_n}$ where $\theta$ is an angle along the circle of
radius $L$. In this variables the  Hamiltonian eq.(\ref{Hamiltonian}) is  given by 
\begin{equation}\label{e1}
H=
%\left(\frac{2\pi}{L}\right)^2\bigg[
\frac{1}{2}\sum_{j=1}^{2N}(z_j\partial_j)^2+\sum_{i \neq j}^{2N}\frac{\lambda(\lambda-1)}{|z_i-z_j|^2}
%\bigg]
\,.
\end{equation}
The ground state of the Hamiltonian wave function of  (\ref{e1}) can be found exactly
\begin{equation}
\nolinebreak{\Psi_0=\left(\prod_{i=1}^{2N}z_i\right)^{-\lambda(2N-1)/2}|\Delta|^{\lambda-1}\Delta}\,, 
\end{equation}
where 
\begin{equation}
\Delta=\prod_{i<j}^{2N}(z_i-z_j)
\end{equation}
is Vandermonde determinant.
Excited states are given by
\begin{equation}
\Psi_{\kappa}=\Psi_0 J_\kappa, 
\end{equation}
where Jack polynomials  $J_\kappa$ are parameterized by partition $\kappa$. 
The problem had been reduced to the properties of the new bosonic  Hamiltonian 
\begin{equation}
H_{\rm B}=\Psi_0^{-1}H\Psi_0
\end{equation}
that acts  in the Hilbert space of symmetric wave functions.
It is given by
\begin{equation}\label{e3}
H_{\rm B}=\sum_{i=1}^{2N} D_i^2+\lambda\sum_{i<j}^{2N}\frac{z_i+z_j}{z_i-z_j}(D_i-D_j)\, ,
\end{equation}
where $D_i=z_i\partial_i$. 
Aiming  for a second quantization one
defines so called collective variables
\begin{eqnarray}&&
\nolinebreak{p(\theta)=\sum_{i=1}^{2N}\delta(\theta-\theta_i), \,\,\,
p_k=\int_0^{2\pi}d\theta e^{ik\theta}p(\theta)},
\nonumber \\&&
p(z)=\sum_{k=-\infty}^{\infty}z^{k-1}p_{-k} \, .
%p(\theta)=\frac{1}{2\pi}\sum_{k=-\infty}^{\infty}e^{ik\theta}p_{-k}
\end{eqnarray}
In terms of collective variables the bosonic Hamiltonian can be rewritten as\cite{Awata}
\begin{eqnarray}&&
\label{e4}
\hspace{-0.5cm}H_{\rm B}=\!
\frac{1}{2}\!\!\!\sum_{m,n=-N}^Nmnp_{n+m}\frac{\partial^2}{\partial
p_n\partial p_m}\!+\!(1-\!\lambda)\!\sum_{n=-N}^N\!
n^2p_n\frac{\partial}{\partial p_n}\!+\nonumber \\&&
\hspace{-.5cm}\frac{\lambda}{2}\sum_{m=0}^{N-1}\sum_{n=1}^{N-m}(n+m)\bigg[p_np_m\frac{\partial}{\partial
p_{n+m}}+p_{-n}p_{-m}\frac{\partial}{\partial p_{-n-m}}\bigg].
\end{eqnarray}
So far the transformation have been exact. 
Passing  to the hydrodynamic limit ($N \to \infty$, $\to \infty$ $2N/L \to \bar{\rho}$
one arrives to eq.(\ref{e5}); $x$ is a coordinate along the circle
($x=\frac{L}{2\pi}\theta$), the linear density
$\rho(x)=\frac{2\pi}{L}\rho(\theta)$.
The modes
of velocity operator are defined as
\begin{equation}
v_n=2\pi\left(-n\frac{\partial}{\partial p_{-n}}+\frac{1}{2}p_n{\rm
sgn(n)}\right). 
\end{equation}
It is easy to see that these definitions
are consistent with a standard commutation \cite{Landau}
\begin{equation}
[v(x),\rho(y)]=-i\delta'(x-y)\, .
\end{equation} 

\end{document}